# The Flex-Pendulum -- Basis for an Improved Timepiece

Randall D. Peters
Department of Physics
Mercer University
Macon, Georgia

ABSTRACT
A compound pendulum of simple geometry was built from a lightweight rod to which a pair of masses are clamped, one above and the other below the axis of rotation. By making the position of the upper mass variable, it was found that the characteristics of this pendulum can be altered dramatically.  A theoretical model of the instrument is presented, which is in reasonable agreement with experimental observations.  From the model it is shown that isochronism of the flex-pendulum can be significantly greater than that of the simple pendulum

INTRODUCTION
It is well known, that early in the history of the pendulum, Huygens invented a method to reduce the effect on the period, due to the nonlinear influence of gravitational restoration.  In his technique, the length of the pendulum was made to be a function of position, by means of cycloidal constraints.
The present studies show that the motion of a compound pendulum can be significantly influenced by nonlinearity associated with elastic flexure.  Although the model presented was first used to describe energy dissipation [1], it also predicts interesting non-dissipative behavior.  For example, improved isochronism can be realized by carefully tailoring the flexure of the rod.

SYSTEM
Photographs and a detailed description of the instrument and its sensor are provided in [2].

**Theoretical Model**
The pendulum was modelled as two point masses separated from each other on a massless rod by a  distance 2L, as shown in Fig. 1.

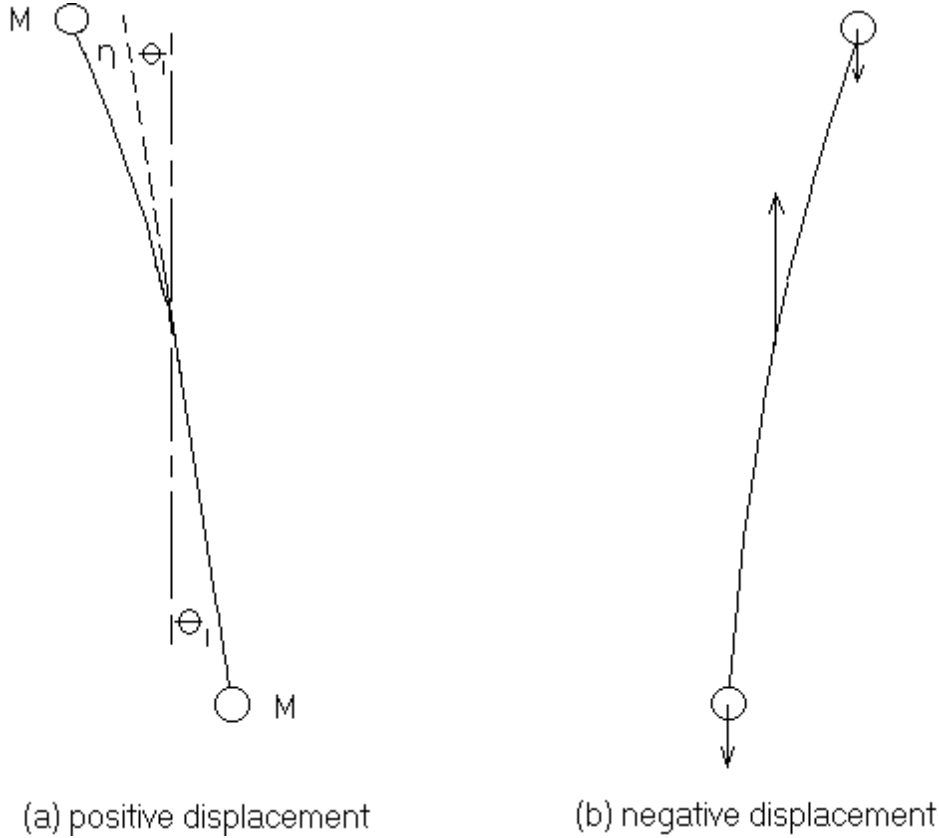

(a) positive displacement    (b) negative displacement

**Figure 1. Illustration of flexure in the compound pendulum.**

With the center of mass located a distance ΔL below the geometric center of the instrument, the equation of motion is

$$\ddot{\theta}_1 + \frac{g}{2L}(1+\frac{\Delta L}{L})\sin\theta_1 - \frac{g}{2L}(1-\frac{\Delta L}{L})\sin\theta_2 = 0 \qquad . \qquad (1)$$

where

$$\theta_2 = \theta_1 + \eta \quad , \quad \eta = c(\theta_1\cos\delta - \frac{\dot{\theta}_1}{\omega}\sin\delta) \quad , \quad \omega = \sqrt{g\frac{\Delta L}{L^2}} \qquad . \qquad (2)$$

Here, the constant c, which is quintessential to present arguments, is related to the elastic constants of the rod supporting the masses. The angular frequency specified in Eq. 2 is valid for c = 1, in the limit as the maximum value of $\theta_1 \ll 1$; i.e., small amplitudes of the motion. As first developed in [1], equations 1 and 2 were concerned with small amplitudes of the motion, so that the sine of the angle was replaced by the angle. As such, they were then used to build the modified Coulomb damping model of hysteretic (internal friction) developed there.

    For the present work, we are not primarily concerned with the damping constant δ. We will be focusing on the elastic properties of the pendulum as determined by c. We will see that the use of c >1 plus nonlinear (sine) terms can result in diverse pendulum behavior. For c = 1, usual simple pendulum behavior is generated; which at low levels corresponds to simple

harmonic motion. For large enough c, otherwise near-harmonic motion is converted to that of the Duffing oscillator. Before arriving at a place where the single-well potential converts to the double-well of the duffing oscillator, elasticity is cause for an amplitude dependence to the period that is oppositely trended to that of the simple pendulum. In principle then, a proper selection for the value of c, relative to other pendulum parameters, could result in an improved timepiece.

In spite of its ingenuity, Huygens' technique did not apparently find wide application among clockmakers. For whatever similar or dissimilar reasons that may exist, the present method is also not expected to be significant to precision clock technology, since there the pendulum has been replaced by quartz and atomic oscillators.

Consideration of the method is important for a different reason. As shown in [2], the long-period compound pendulum is ideally suited to the study of largely unknown, but important internal friction processes--ones that influence the motion of every pendulum. A complete understanding of hysteretic friction requires a complete understanding of the non-dissipative restoration properties of the instrument used to study it. It is felt that this pendulum is a useful pedagogical tool to increase our awareness of the importance of nonlinear damping.

**Simulation Results -- Emphasis on Elastic properties**

Because the equations are nonlinear, we resort to numerical integration to investigate the properties of this pendulum. To do so, the equations were first rewritten as a coupled pair of first order expressions, as is customary because roundoff errors are then reduced.

Shown in Fig. 4 is the trivial case for c = 1.

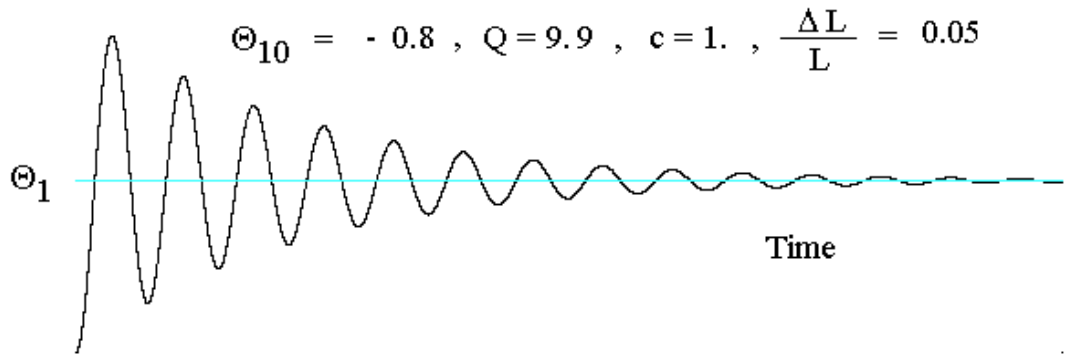

**Figure 2. Example of oscillation for an assumed rigid pendulum**.

In the limit of small motion, the period for the Fig. 2 case would be 6.347 s. It is initially larger than this by 4% (for the initial amplitude of 0.8 rad).

For all of the theoretical graphs provided, damping is provided by a simpler (though less realistic) means than what is indicated in Eq. 2. Damping is accomplished in the common manner, by multiplying by a constant (0.1) the first derivative of $\theta_1$ (angular velocity) where it occurs in the equation of motion. For the total pendulum length of $2L = 1$ m, the Q of the pendulum in the small amplitude limit, with c = 1, is therefore given by [1960 $\Delta L/L]^{1/2}$. (The reader less familiar with the use of Q instead of '$\beta$' in the equation of the simple harmonic oscillator is referred to [2]. )

As the constant c is increased above 1 of the rigid case, there are interesting consequences. First of all, flexure causes 'softening' of the pendulum at all frequencies; i.e., the period of the pendulum is generally lengthened. Additionally, however (and most interesting) there is an amplitude dependence of the softening, that will be discussed later in the context of improving isochronism. The softening effect is illustrated in Fig. 3, which shows period increase for $c = 1.02$ as compared to the rigid pendulum. For the indicated parameters, a 2% flexure has caused a 12% reduction in the frequency.

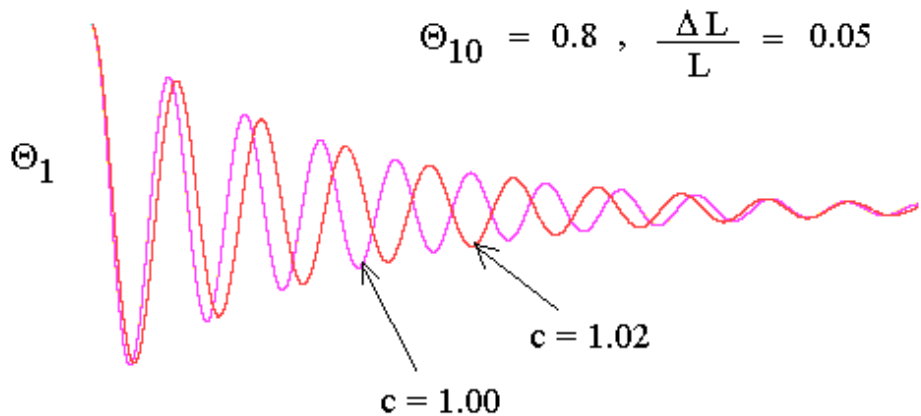

**Figure 3. Illustration of 'softening' as the result of flexure.**

Notice how the initial in-phase condition is quickly lost -- phase reversal after 3 to 4 cycles, and then after 8.5 cycles the two oscillations are back in phase again.
We now look at the Duffing extreme of large c, illustrated in Fig. 4.

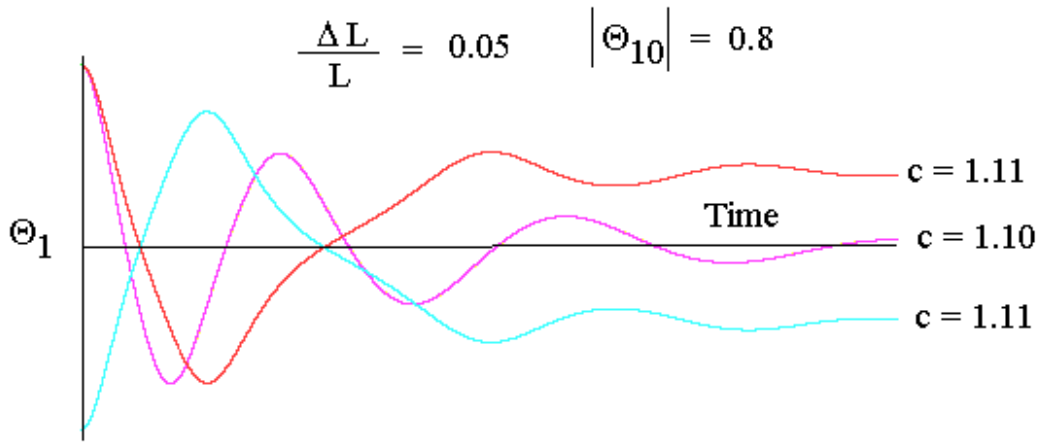

**Figure 4. An increase of c from 1.10 to 1.11 causes the single-well, anharmonic potential to convert to a double-well of the Duffing type.**

Notice from Fig. 4, that the case of c = 1.11, with initial positive displacement, winds up in a different potential well than that of initial negative displacement.

The sensitive dependence on initial (SDI) conditions that is characteristic of the Duffing oscillator is illustrated in Fig. 5, where the initial displacements are almost the same, but the final outcomes are largely different.

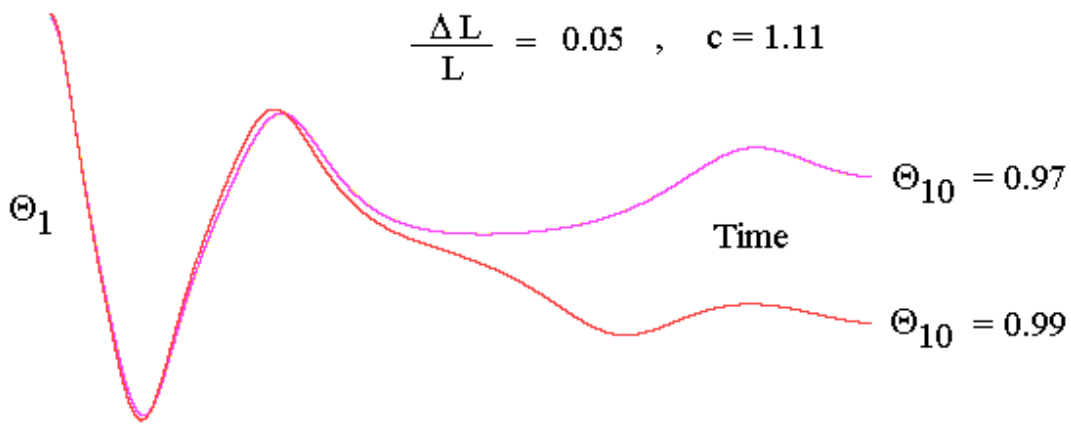

**Figure 5. Illustration of SDI that is characteristic of the Duffing oscillator**.

It is instructive to consider the manner in which the potential well changes shape as a function of c, in the vicinity of the critical value (1.106), where conversion to the Duffing oscillator occurs. This is shown in Fig. 6.

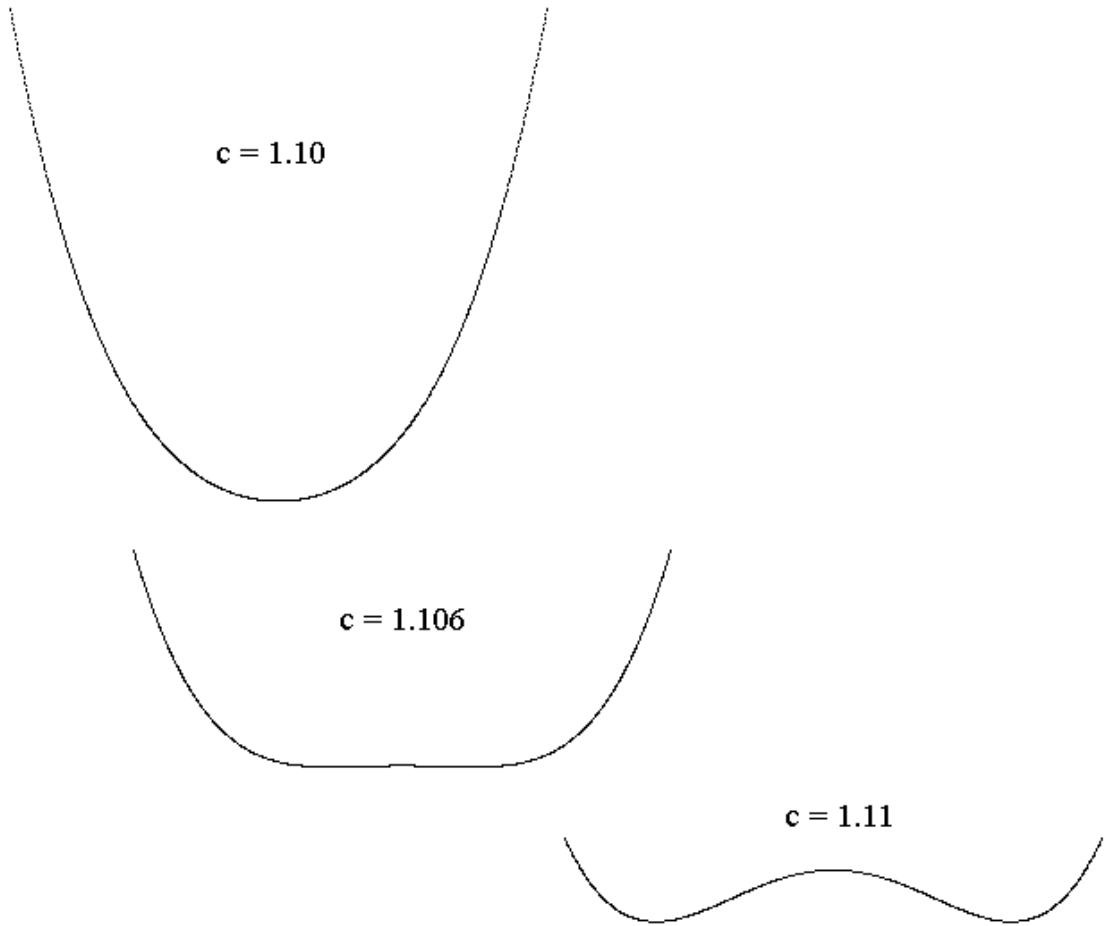

**Figure 6. Variations in shape of the potential energy as a function of the flex-constant c.**

**Tailoring for Improved Isochronism**

It was assumed that partial compensation for the change of period with amplitude of the simple pendulum might be realized for some configurations of the flex-pendulum--if its softening character should have the 'proper' amplitude dependence. Fig. 7 shows that the direction of the change is in support of the premise. Relative to its value at the start of the decay, the period is seen to have lengthened by a factor of about 2 by the end of the record.

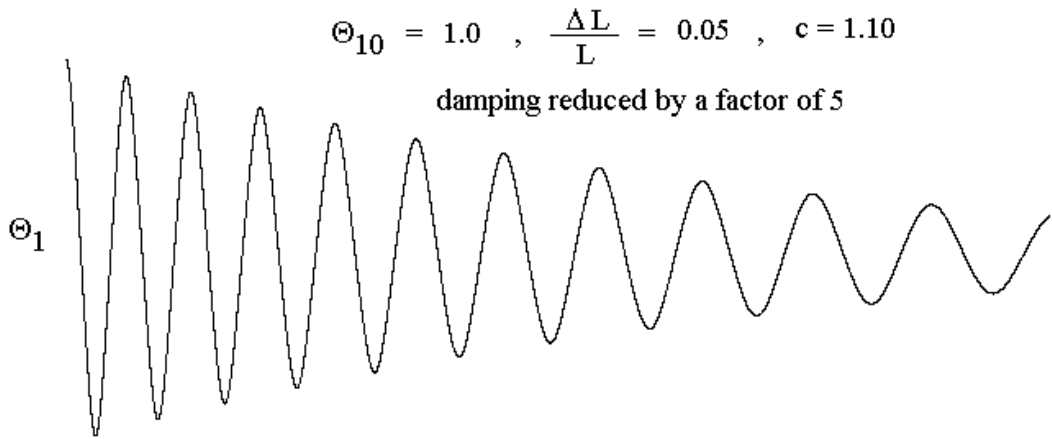

**Figure 7. Flexure causing an amplitude trend for the period that is opposite to that of the simple pendulum.**

From Fig. 7 it was noted that flexures smaller than 10% would most likely be required of any case of serious compensation. An analytical estimate for the optimum value of c was not deemed feasible; so a trial and error search for compensation was initiated, using using the computer.

A fixed frequency reference harmonic oscillation was generated, by integrating the simple harmonic oscillator equation of motion. The frequency of this reference was chosen to match the pendulum frequency, for a given value of c, when the pendulum amplitude was set to a small value. For all comparisons, the damping was turned off, and the simulated pendulum oscillation was used with the reference oscillation to produce a Lissajous figure on the monitor of the computer.

For the family of results associated with a given value of c, the following procedure was then used. With small amplitude synchronization confirmed, the amplitude of the pendulum's oscillation was increased in steps upward. From the time required for a beat between the pendulum and the reference (observed by periodically refreshing the figure on the monitor), the frequency difference was calculated. This was done by multiplying the number of cycles per display (typically n = 19) by the number of refreshes required for a complete beat (N). Thus the percentage deviation in the period of the pendulum from the reference was obtained by

$$\% \text{ dev.} = \frac{100}{nN}$$

Results for one value of c are are indicated in Fig. 8.

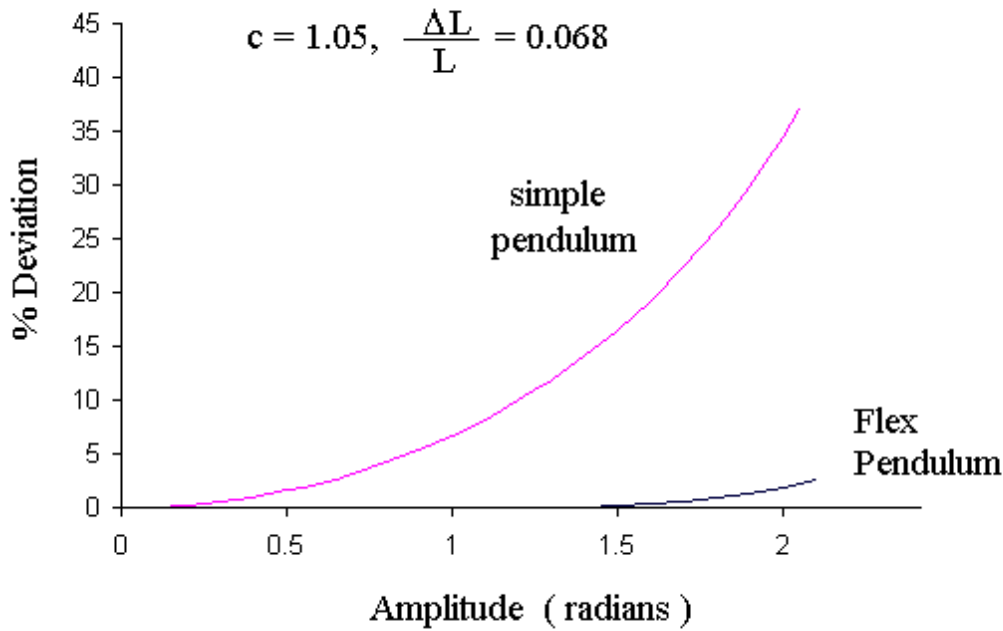

**Figure 8. Example of improved isochronism for the parameters indicated**.

The reference period of oscillation for the case of Fig. 8 is 5.442 s. At an amplitude of 1.5 rad the period of the uncompensated pendulum would increase about 15 %. The increase for the flex-compensated pendulum at this same amplitude is more than two orders of magnitude smaller. For this specifically tailored flex-pendulum--we see that, for motion limited to $\pm\pi/2$, any dependence on amplitude of the period is virtually undetectable!.

**SUPPORTING OBSERVATIONS FROM EXPERIMENT**

      A full experimental validation of the proposed method to build a highly compensated flex-pendulum has not yet been attempted. Just as the theory used to generate Fig. 8 has required numerous trial and error computer runs, so building an actual pendulum with the spectacular properties of Fig. 8, could be very time consuming under the best of conditions. With less favorable conditions, fabrication of the mechanical structure could require expertise that is not possessed by, or readily available to the author.
      What was done experimentally is to show that the pendulum behaves in a manner consistent with some of the figures that are preliminary to the final result of Fig. 8. For each of the cases that follows, two lead masses were used--one (993 g) adjustable-in-position, and clamped above the axis, and the other (711 g) fixed-position, and clamped at the bottom of the pendulum rod (40cm below the axis).

**Illustration of the Need for Compensation**

To illustrate the usual amplitude dependence of the pendulum, the upper mass was located 17 cm above the axis to generate the data shown in Figure 9. Extra air damping was provided (to show the effect in a shorter time) by taping a floppy disc to the rod just above the bottom weight.

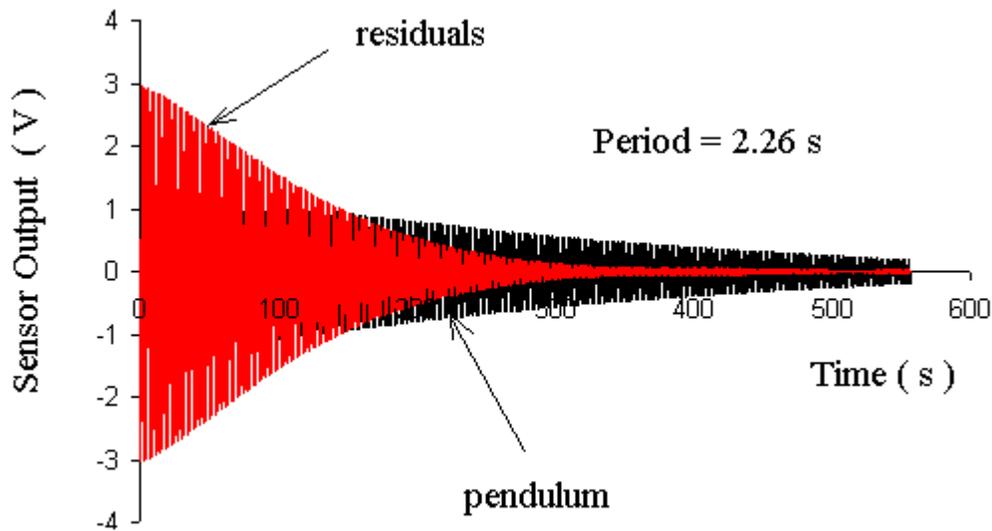

**Figure 10. Illustration of non-isochronism in the near-rigid pendulum**.

Displayed in Fig. 10 are two traces: (i) the pendulum decay (black), and (ii) the difference (residuals) between the pendulum motion and a damped sinusoid that was computer fitted to the data (red).  The frequency and phase of the computer generated fit were selected to give small residuals in the latter part of the decay, where the motion is more nearly isocrhronous.  At the start of the record, the pendulum and the reference sinusoid are 180 degrees out of phase, because of the large amplitude (0.39 rad).  The change in phase over the first 300 s is consistent with the uncompensated (simple) pendulum curve of Fig. 8.

**Duffing Case**

Shown in Fig. 11 is a Duffing case, which was obtained by moving the upper mass to a point 27 cm above the axis.

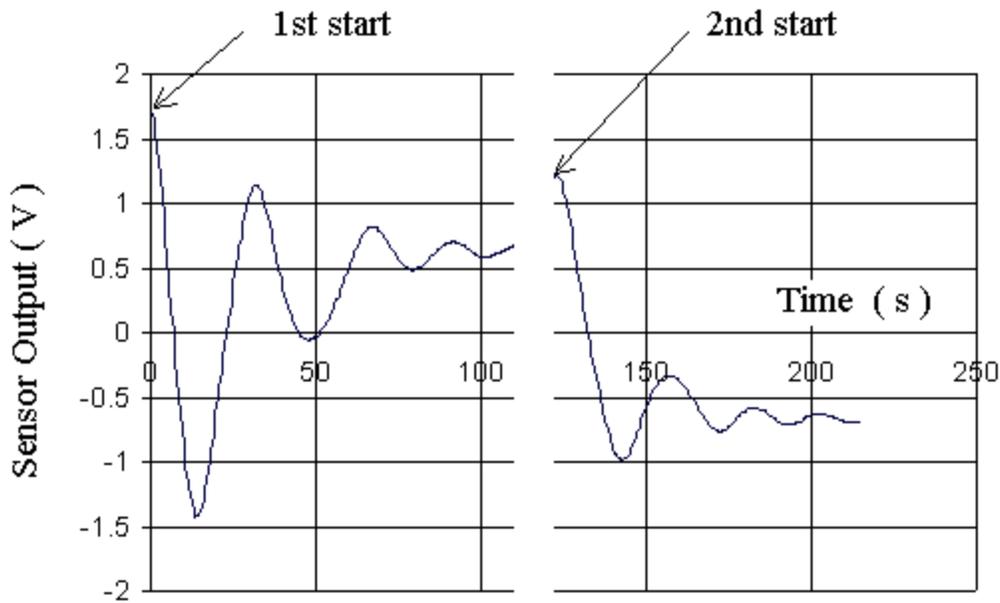

**Figure 11. An example of different initial displacements yielding different final equilibrium positions.**

The Duffing character is illustrated by starting the pendulum at different initial displacements, the velocity there being zero each time. For the sensor calibration constant of 3.85 V, the first start was at 0.44 rad (25 degrees); and the second was at 0.32 (19 degrees). The ultimate final 'attractor' is different for the two cases in a manner consistent with Fig. 5. Actual demonstration of SDI with the pendulum is more difficult experimentally than theoretically because of creep-- which at these long periods is not ignorable.

      In Fig. 12 it is seen that the pendulum can be configured so that the period lengthens significantly as the amplitude decreases. It was obtained by lowering the upper mass slightly as compared to the Duffing case.

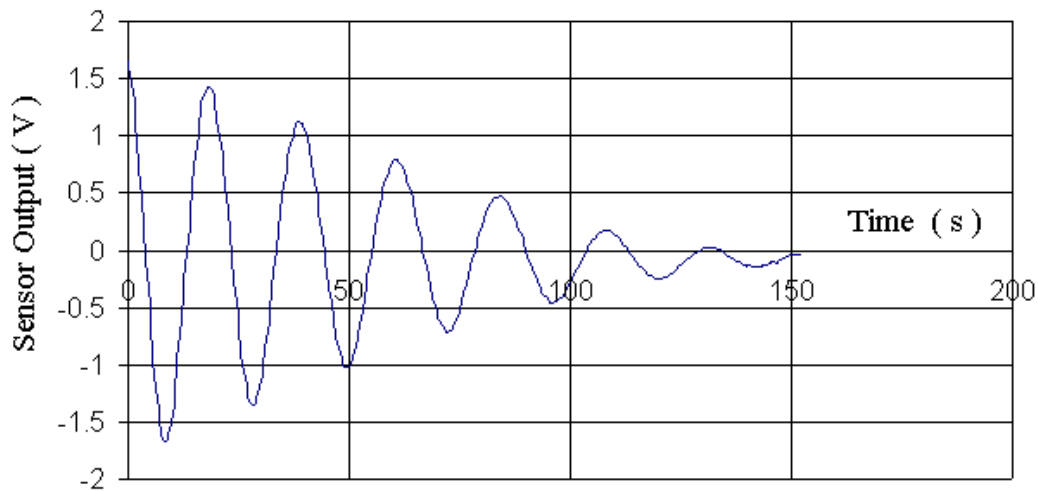

**Fig. 12. Illustration of period vs amplitude variation in the flex-pendulum. The trend is opposite to that of a rigid pendulum.**

## CONCLUSIONS

Although the experimental results only demonstrate feasibility, they are important because of their good agreement with theory for the cases considered. It is therefore reasonable to expect that the flex-pendulum can be configured to be much more isochronous than the simple pendulum.